\documentclass[journal=nalefd,manuscript=letter]{achemso}
\usepackage[T1]{fontenc}       
\usepackage{amsmath}
\usepackage{amsfonts}
\usepackage{amssymb}
\usepackage{graphicx}
\usepackage{pifont}
\DeclareFontFamily{U}{astrosym}{}
\DeclareFontShape{U}{astrosym}{m}{n}{<8> astrosym}{}
\usepackage{achemso}
\setkeys{acs}{maxauthors = 0}
\usepackage{caption}
\usepackage{subcaption}
\usepackage{tikz}
\usepackage{bbold}
\usepackage{stackengine}
\usepackage{bbm}

\author{D. Weckbecker}
\affiliation{Lehrstuhl f\"ur Theoretische Festk\"orperphysik, Staudtstr. 7-B2, 91058 Erlangen, Germany}
\author{M. Fleischmann}
\affiliation{Lehrstuhl f\"ur Theoretische Festk\"orperphysik, Staudtstr. 7-B2, 91058 Erlangen, Germany}
\author{R. Gupta}
\affiliation{Lehrstuhl f\"ur Theoretische Festk\"orperphysik, Staudtstr. 7-B2, 91058 Erlangen, Germany}
\author{W. Landgraf}
\affiliation{Lehrstuhl f\"ur Theoretische Festk\"orperphysik, Staudtstr. 7-B2, 91058 Erlangen, Germany}
\author{S. Leitherer}
\affiliation{Department of Micro- and Nanotechnology, Technical University of Denmark, \O{}rsteds Plads, Building 345C, 2800 Kgs. Lyngby, Denmark}
\author{O. Pankratov}
\affiliation{Lehrstuhl f\"ur Theoretische Festk\"orperphysik, Staudtstr. 7-B2, 91058 Erlangen, Germany}
\author{S. Sharma}
\affiliation{Max-Born Institute for Nonlinear Optics and Short Pulse Spectroscopy, Max-Born-Strasse 2A, 12489 Berlin, Germany}
\author{V. Meded}
\affiliation{Karlsruhe Institute of Technology, Institute of Nanotechnology, Hermann-von-Helmholtz-Platz 1, 76344 Eggenstein-Leopoldshafen, Germany}
\author{S. Shallcross}
\affiliation{Lehrstuhl f\"ur Theoretische Festk\"orperphysik, Staudtstr. 7-B2, 91058 Erlangen, Germany}
\email{sam.shallcross@fau.de}

\title{Moir\'e ordered current loops in the graphene twist bilayer}
\date{\today}

\begin{tocentry}
\includegraphics[width=1.0\textwidth]{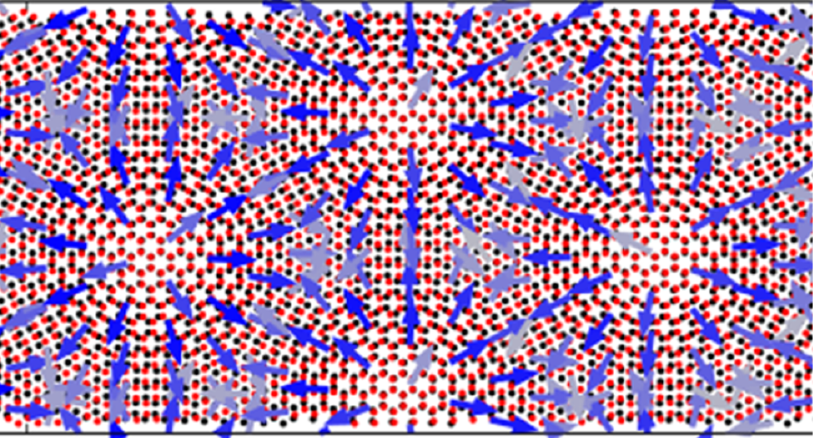}
\end{tocentry}

\begin{document}

\begin{abstract}

While a typical material exhibits field induced currents only at the boundary, a uniform out-of-plane magnetic field applied to two mutually rotated layers of graphene is shown to result in an ordered array of permanent current loops throughout the material. Each current loop consists of an interlayer current flowing through the open AA stacked regions of the moir\'e created by rotation, which then flows back through the neighboring AB regions to form a circuit, with significant current strength even at small fields.
Similar moir\'e ordered arrays of current loops are also shown to exist in non-equilibrium transport states, where they manifest as current back flowing against the applied bias in the device. Such current loops thus represent an intrinsic feature of the twist bilayer in conditions of broken time reversal symmetry, and exist both as a low field imprint of the moir\'e lattice on Landau physics, and as measurable moir\'e scale current configurations in transport states.

\end{abstract}

\maketitle

{\bf Introduction.} For a typical material the magnetic length exceeds by many orders of magnitude the lattice parameter. The material can thus be approximated as homogeneous, leading to the well known non-relativistic ($E_n \propto n+1/2$) or Dirac-Weyl ($E_n \propto \sqrt{n}$) Landau level spectra. This situation may change in a layered two dimensional system, for which a mutual rotation between the layers can result in a long range stacking modulation known as a moir\'e lattice. Such materials have a moir\'e lattice parameter comparable in magnitude to the magnetic length of experimentally achievable magnetic fields and, as a consequence, dramatically different physics in the presence of a magnetic field\cite{Dean2013a,Hunt2013a,pon13}. Most notably this can be seen through the fact that the $B$ or $\sqrt{B}$ field dependence of the standard non-relativistic or Dirac-Weyl spectra goes over to a fractal dependence on the field strength (the so called Hofstadter butterfly\cite{Hofstadter1976}).

The graphene twist bilayer\cite{hass08,shall08,shall08a} is perhaps the most studied example of a two dimensional moir\'e. This system exhibits a remarkable range of electronic structure phenomena, encompassing both graphene like (large angle) and charge confined (small angle) limits\cite{lan13,shall13,shall16}. For small twist angles the moir\'e is comparable to the magnetic length of achievable (1-5~Tesla) experimental fields, and indeed for a $\theta = 0.52^\circ$ twist bilayer signatures of the Hofstadter butterfly were recently observed in experiment\cite{Kim2017}.

In this work we show that in addition to displaying the Hofstadter butterfly, the twist bilayer possesses a low field and intermediate angle pre-Hofstadter phase in which although the Landau spectrum has the $E_n \propto \sqrt{n}$ Dirac-Weyl form of graphene, the wave functions carry a remarkable imprint of the moir\'e in the form of an array of permanent current loops, with one current loop per moir\'e cell. This current loop lattice occurs at significantly weaker fields than those required to observe the Hofstadter butterfly: for example, for a twist angle of $6^\circ$ it is found already at a field of the order of 1 Tesla, while a field of $\approx$118 Tesla would be required to observe the Hofstadter butterfly. While such current loops might be difficult to observe in equilibrium conditions -- although they will certainly impact intercalation of the bilayer -- we show they exist also in transport states, both with and without an external magnetic field. To see this we gate a single layer of graphene and overlay this with a twisted layer. This system exhibits an ordered array of current loops aligned in the bias direction, resulting in remarkable back flowing current against the bias direction in the ungated layer. Such ordered arrays of current loops should thus be observable in transport experiments under ballistic conditions, for instance with scanning probe techniques\cite{baum07,mor15,bour17}.

\begin{figure}[t!] 
  \begin{center} 
  \includegraphics[width=0.4\linewidth]{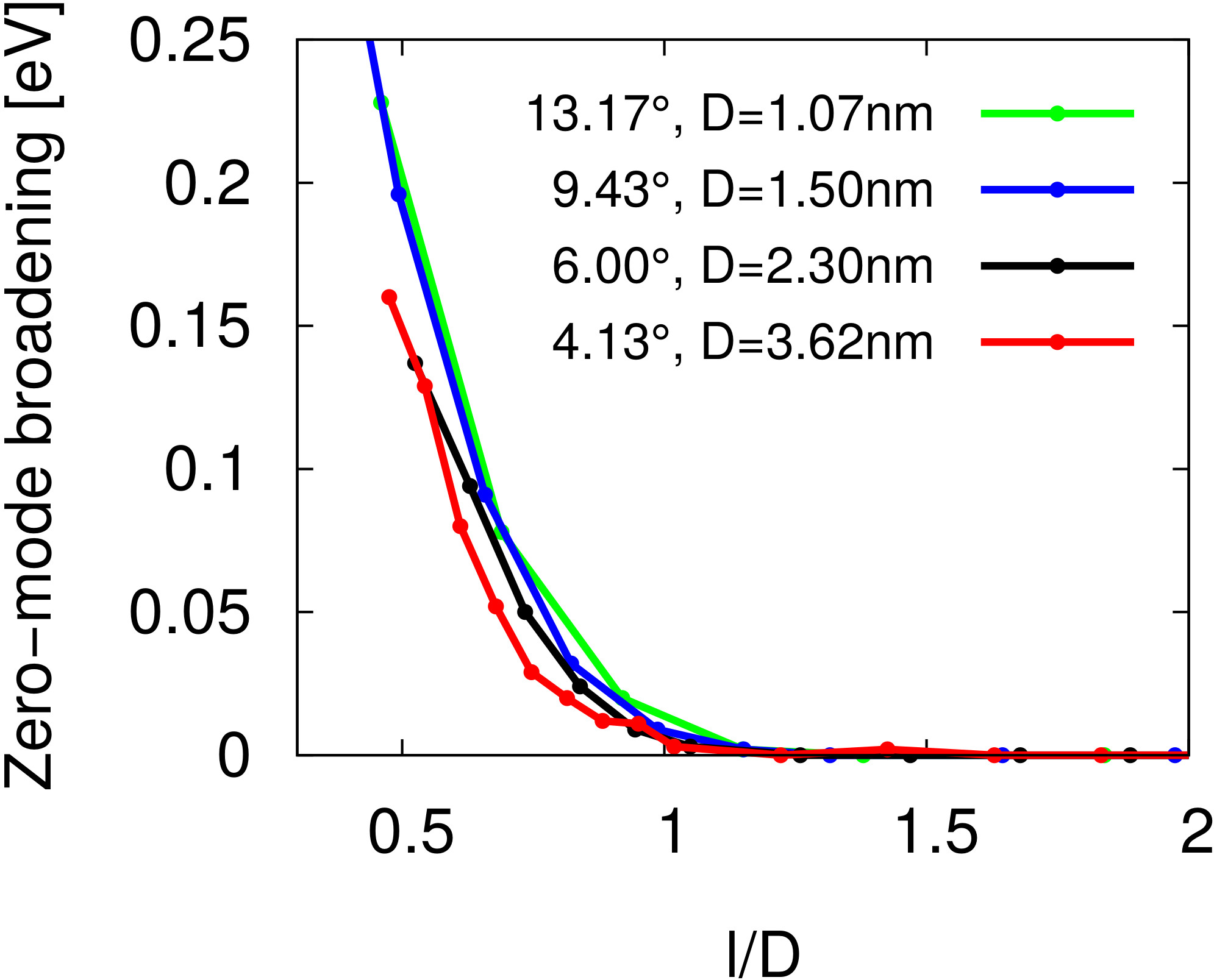}
  \caption{\emph{Broadening of the zero mode Landau level as a function of the ratio between the magnetic length $l$ and the moir\'e periodicity $D$}. The zero mode begins to significantly broaden only when $l$ becomes smaller than $D$, a universal behavior found for all twist angles in the moir\'e regime of $\theta \le 15^\circ$. This broadening of the zero mode can be taken as establishing a critical field strength for the onset of the Hofstadter butterfly.}
  \label{Zbroad}
  \end{center}
\end{figure}

{\bf Results.} In order to address physics in the pre-Hofstadter regime, we must first establish a criteria for the critical field strength at which the Hofstadter butterfly is first seen. Such a criteria has been suggested by Moon and Koshino\cite{moon12} to be $l_B < D$, with $l_B=\sqrt{\hbar/(eB)}$ the magnetic length and $D=a/(2\sin\theta/2)$ the moir\'e length ($a$ is the lattice constant of graphene). From the semi-classical point of view this is an intuitive result, as once the magnetic length exceeds the underlying moir\'e periodicity it is effectively averaged out over one cyclotron orbit, and thus one expects that quasiparticles will feel only a structureless average of the moir\'e. We will therefore first numerically investigate whether this criteria represents a useful marker of the pre-Hofstadter regime.

\begin{figure}[ht!] 
    \begin{subfigure}[c]{0.50\textwidth}
	\topinset{}{\topinset{}{\includegraphics[width=0.95\linewidth]{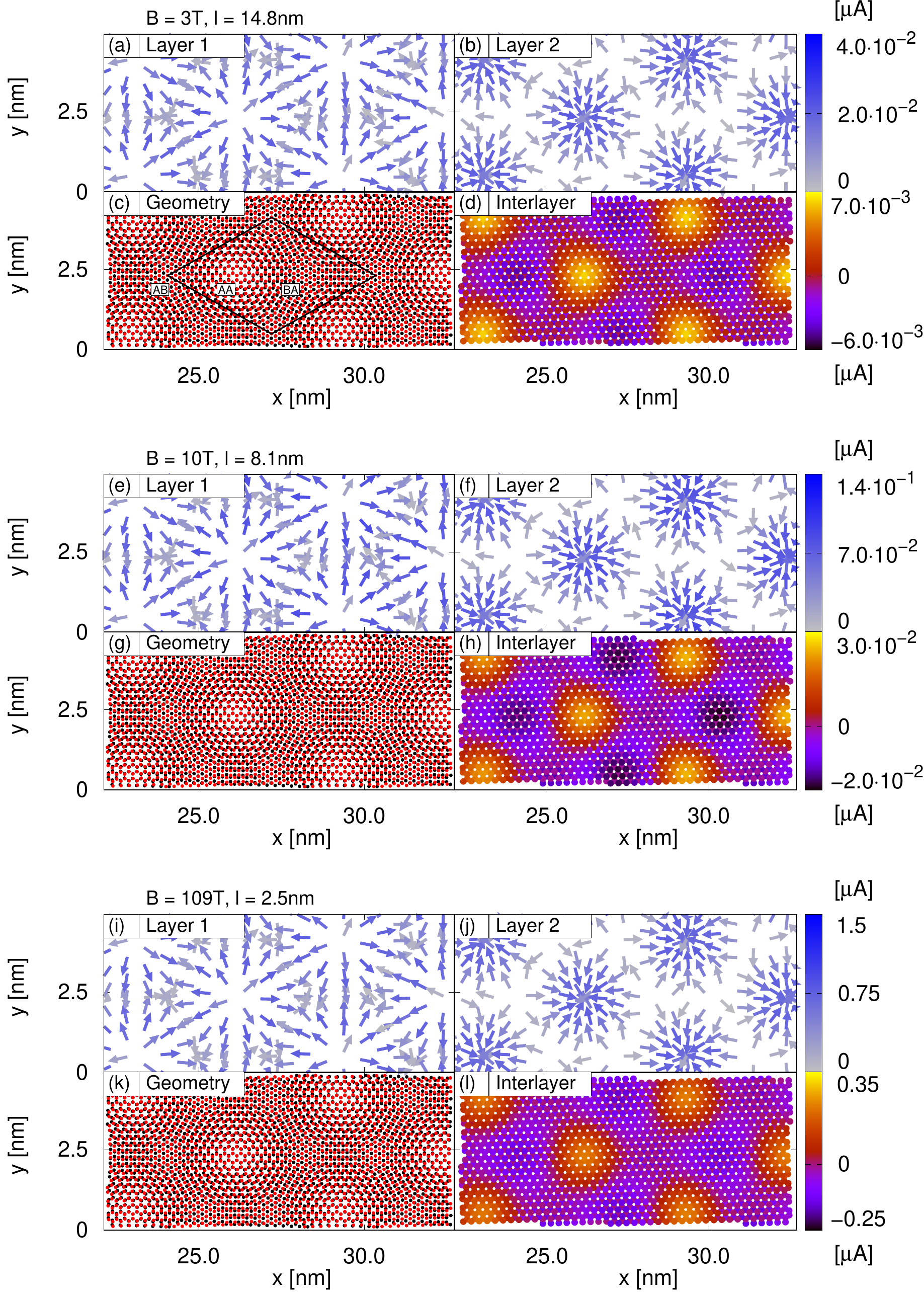}}{-0.5cm}{-3.5cm}}{0cm}{0cm}
    \end{subfigure}\hfill
    \begin{subfigure}[c]{0.50\textwidth}
    	\topinset{}{\topinset{}{\includegraphics[width=0.95\linewidth]{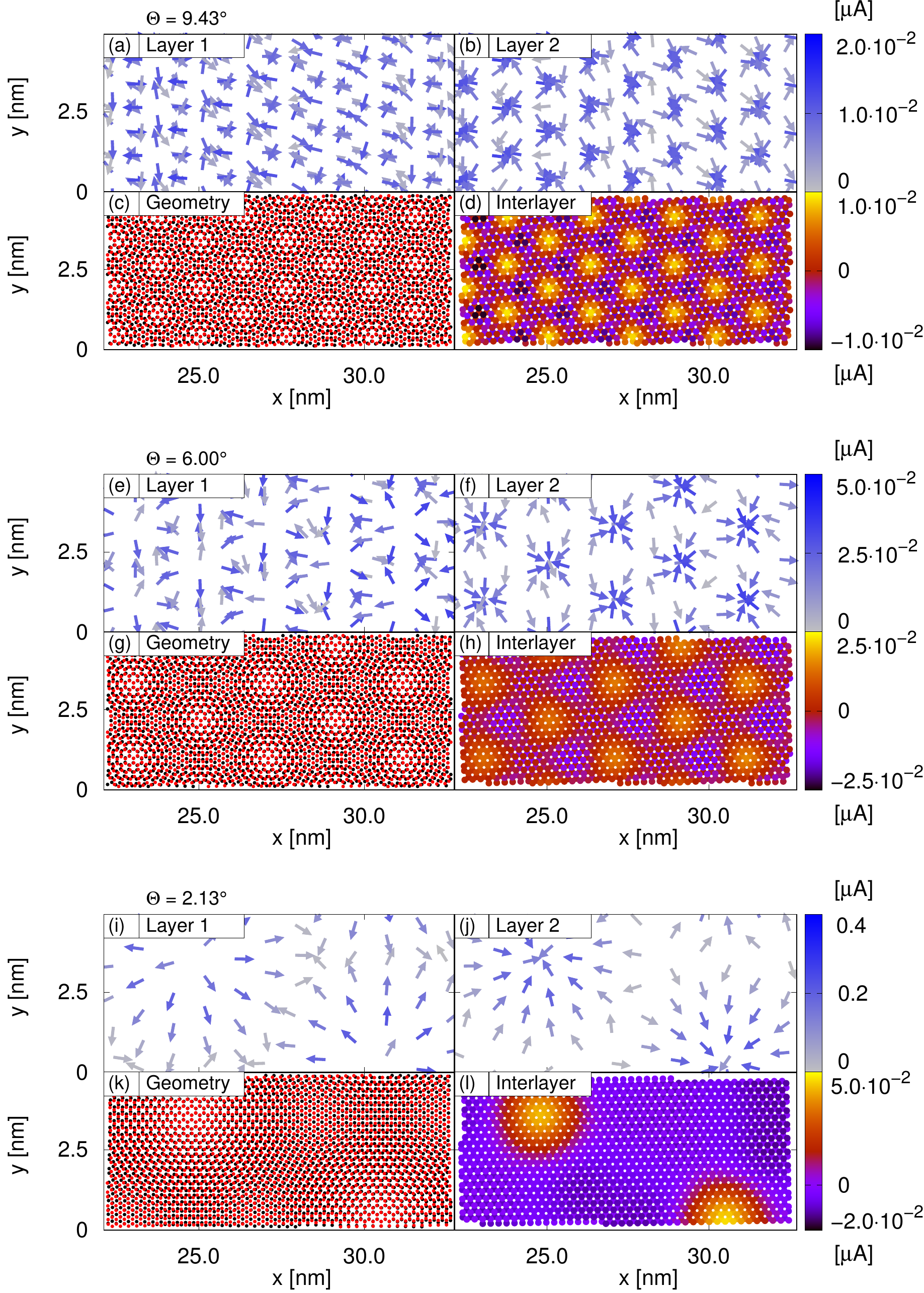}}{-0.5cm}{-3.5cm}}{0cm}{0cm}
    \end{subfigure}
   \caption{\emph{Permanent current loops in a graphene twist bilayer for (left panel) different magnetic fields with the twist angle fixed at $\theta = 3.89^\circ$, and (right panel) different twist angles with the magnetic field fixed at 10 Tesla}: Panels (a)-(b) present in-plane currents summed over the zero mode of the nanoribbon in an out-of-plane magnetic field of $B = 3~\mathrm{T}$ (the filling factor of the zero mode is taken to be 1, although similar results are seen for other filling factors). The arrows indicate the direction of the in-plane component of the current in the first (a) and second (b) layers, while the color of the arrow indicates the strength of the currents. (c) Corresponding geometry of the twist bilayer, with black and red dots indicating carbon atoms from the first and second layer respectively. The moir\'e lattice contains all possible stacking types within a unit cell indicated by the full black line, with the three high symmetry cases labeled by AA, AB, and BA. (d) Interlayer currents integrated over the zero mode; the colors refer to the strength of the perpendicular current going from the first to the second layer. The remaining blocks show the same properties
   \label{Bcur}, but for different fields and/or twist angles.}
\end{figure}

For this purpose the width of the zero mode Landau level provides a useful guide as (i) at the onset of the Hofstadter butterfly the degeneracy of this mode will be broken, and thus will acquire a finite width and, (ii), the zero mode is the last of the Landau levels to be destroyed by the clustering of van Hove singularities towards the Dirac point as $\theta\to 0$\cite{shall13,shall16}. We therefore consider four different twist angles in the moir\'e regime of $\theta < 15^\circ$ and plot the zero mode width as a function of applied field. As shown in Fig.~\ref{Zbroad}, for all four twist angles the onset of zero mode broadening is always at $l/D \approx 1$, confirming the intuitive criteria proposed in Ref.~\cite{moon12}. Note that the zero mode broadening curves are seen to be, to a good approximation, a function only of the ratio $l/D$ suggesting that the lattice constant $a$ no longer plays an important role at the large $D$ found at these twist angles.

The current density of the ground state zero mode wave function, as we now show, turns out to be profoundly sensitive to the moir\'e lattice even in the pre-Hofstadter regime of $l_B >> D$. In Fig.~\ref{Bcur} we present the current density obtained by summing over all states in the zero mode of a twist bilayer nanoribbon with $\theta = 3.89^\circ$. Shown are the current densities for three different fields: $B=3~\mathrm{T}$ [panels (a-d)], $B=10~\mathrm{T}$ [panels (e-h)], and $B=109 ~\mathrm{T}$ [panels (i-l)]. The band structures and corresponding density of states for these systems can be found in SI; one can note that the moir\'e induced van Hove singularity is at $\sim\pm0.3$eV, greater than $\hbar \omega_c$ except for very large fields. The first two of these fields have $l_B/D$ ratios of 4.14 and 2.23 respectively, and exhibit clear $\sqrt{n}$ Dirac-Weyl Landau spectra (see SI). In contrast, the $B=109 ~\mathrm{T}$ field has $l_B/D = 0.69$ and a zero mode broadening greater than $\hbar \omega_c$ (see SI): this is clearly in the Hofstadter regime. In all cases, however, the current density of the zero mode is qualitatively the same and exists throughout the bulk of the material. Remarkably, therefore, the Landau wave functions exhibit a current loop lattice at magnetic fields for which the Landau spectra shows no trace of the presence of the moir\'e.

\begin{figure}[t!] 
  \begin{center} 
    \begin{subfigure}[c]{0.85\textwidth}
        \centering
	\includegraphics[width=\linewidth]{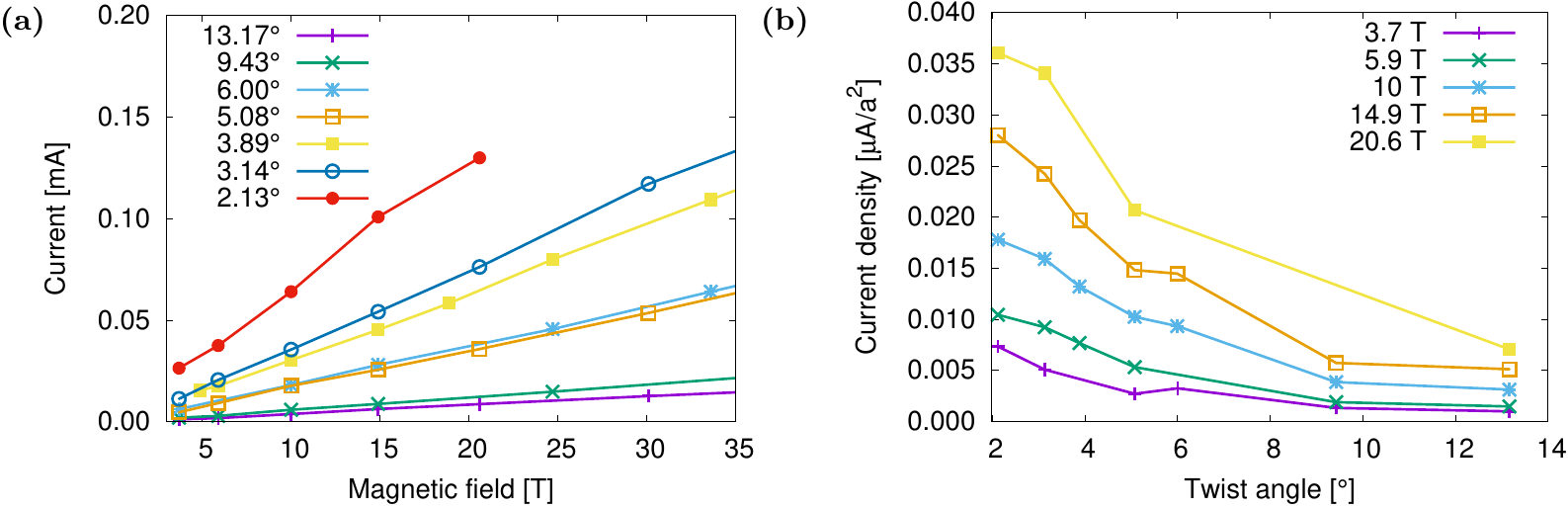}
    \end{subfigure}
  \caption{\emph{Strength of the current loops}. (a) Average interlayer current through the AA region of the moir\'e lattice of the zero mode Landau level (filling factor 1), plotted as a function of the magnetic field for several twist angles. The legend indicates the twist angle in each case. The deviations for linear behavior seen for the smaller twist angles occur for fields for which the magnetic length is smaller than the moir\'e length. (b) The current density plotted as a function of the twist angle for a range of magnetic field strengths. Decreasing the twist angle leads to a monotonic increase in the strength the interlayer current density.}
  \label{Jstrength}
  \end{center} 
\end{figure}

We now characterize more closely the current density. This consists of permanent current loops in the bulk of the material in which an interlayer charge flows through the AA spots, and back through the neighboring AB spots of the lattice in the opposite direction. As may be seen from panels (d), (h), and (l), the interlayer current is almost zero in the intermediate stacking regions with, however, an in-plane current such that a closed loop is formed between the ``source'' and ``sink'' regions of interlayer current in each layer. The topology of the current flow thus has the structure of a ``convection cell'' attached to each moir\'e spot of the lattice.  Furthermore this current loop lattice, irrespective of the strength of the field, has a lattice parameter equal to the moir\'e lattice parameter; a fact evident from the similar form of the current density seen in all three panels of Fig.~\ref{Bcur} even though the magnetic field changes by more than an order of magnitude.

Note that although we have shown the current density for a zero mode filling factor of one, a similar structure of the current density is found for any finite filling factor, although for low filling factors ``vacancies'' appear in the loop lattice (see SI). This simply reflects the fact that individual eigenstates all exhibit a similar current density, modulated however by an exponential envelope on the length scale of $l_B$. Summing these over the zero mode then leads to the current shown in Fig.~\ref{Bcur} in which only the presence of the length scale $D$ can be seen.

We now consider the strength of these current loops. As the interlayer current averaged over the bulk of the nanoribbon must obviously be zero, to characterize its strength we instead sum over all currents within a single AA spot of the moir\'e, within which the interlayer current always flows in the direction of the applied field. The field dependence of this AA current grows to a very good approximation linearly in all cases (see Fig.~\ref{Jstrength}(a)), provided $l_B > D$. This arises from the fact that (i) the degeneracy of the zero mode increases linearly with field and, (ii), as noted above, all eigenstates in the zero mode have a similar current density. These net currents are significant in size, of the order of 10-100~$\mu$A, larger by an order of magnitude than the currents typically found in driven nanostructures\cite{Weckbecker2017,Ullmann2015,Hofmeister2014,Motta2012,Benesch2009}. To investigate how the moir\'e impacts the intrinsic current strength in Fig.~\ref{Jstrength}(b) we present the current density of the AA spots, i.e. the total current through the AA region divided by its area, as a function of twist angle for several fields. Interestingly, the current density increases significantly as the twist angle decreases, with for the 5.9 Tesla field strength an almost order of magnitude increase in the current density found on reducing the twist angle from 12$^\circ$ to $2^\circ$.

\begin{figure}[t!]
   \centering
    \begin{subfigure}[c]{0.95\textwidth}
        \centering
	\includegraphics[width=\linewidth]{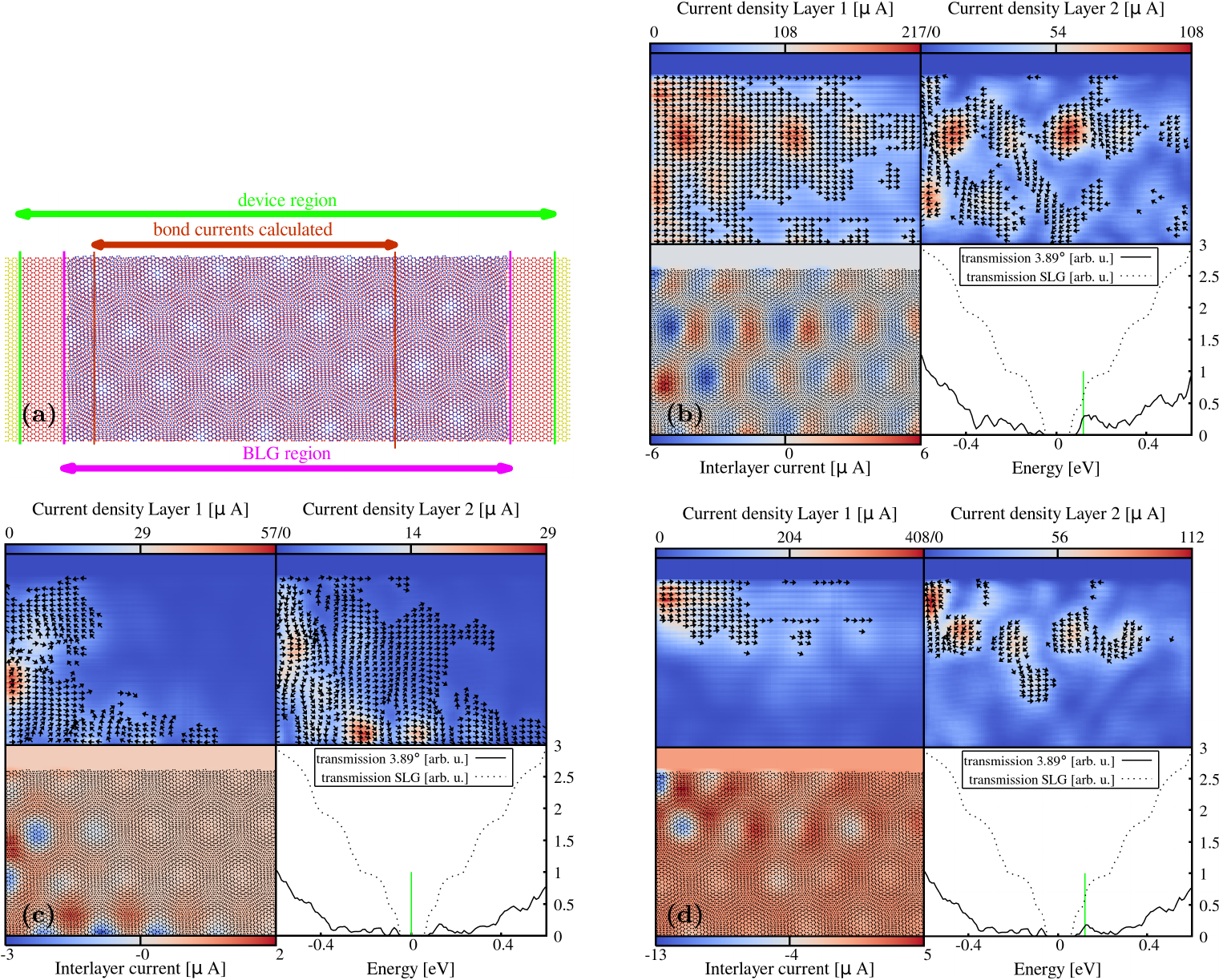}
    \end{subfigure}
  \caption{\emph{Current loops in transport states of the twist nanoribbon}. In panel (a) is shown the transport setup which consists of a graphene nanoribbon junction (of width $l_W \sim 100$~nm) overlaid by a twisted layer ($\theta = 3.89^\circ$) not connected to the leads. Scattering states for this setup are shown in panel (b) for zero external $B$-field, and in panels (c) and (d) for an external $B$-field with $l_W/l_B \sim 4$ ($l_B$ is the magnetic length). In the absence of the magnetic field one finds ordered arrays of current loops that together result in a current flow in the second layer against the transport direction of strength $\sim50$\% of the current in the gated layer, see panel (b). However, in contrast to the ground state at finite field, see Fig.~2, the interlayer current now changes sign on the AA patches of the moir\'e. In the presence of a magnetic field one finds that at energies for which the longitudinal transport is blocked the current loop structure is similar to that found in the ground state (interlayer current alternating sign between the AA and AB/BA patches), while for transmitting states the current configuration more closely resembles that of $B=0$ scattering states.}
  \label{transfig}
\end{figure}

Closed current loops in equilibrium are likely to be difficult to detect experimentally and so we now turn to the question of whether signatures of this unusual ground state exist in transport states, for which we employ a non-equilibrium Green's function technique (see the Methods section). Our transport setup is shown in Fig.~4(a) and consists of a single layer graphene nanoribbon junction (width $\sim100$~nm) overlaid by a rotated layer ($\theta = 3.89^\circ$) not connected to the leads. Current flow in the top layer will therefore arise purely from the interlayer interaction.

In Fig.~4(b-d) we show the local currents in the area of the junction indicated in 
Fig.~4(a) for three representative low energy scattering states. Each of (b-d) display in the topmost panels the in-plane currents in layers 1 and 2, and in the bottom panels the interlayer current density (left hand panel) with the lattice sites displayed as small black dots, and in the right hand panel the left-to-right transmission function of the junction. In this latter panel the vertical line indicates the energy of the scattering state for which the currents are shown.

We first consider transport in the absence of the out-of-plane magnetic field, Fig.~4(b). We note that the transmission function is lowered by the presence of the overlaid twist layer, in agreement with recent finite temperature Boltzmann calculations of the twist bilayer\cite{ray16}. However most interesting is the structure of the electron currents in the overlaid layer which flow against the transport direction, a remarkable situation that arises from the formation of a moir\'e ordered array of interlayer current loops.
Contrasting Figs.~4(b) and 2 one notes that this loop lattice no longer possesses the approximate $C_3$ symmetry of the underlying moir\'e, with the current loops now centered on the AA regions of the lattice. For transport states at higher energies the loop lattice disorders, although strong interlayer currents remain a feature of the transport current.
Thus the ordered array of current loops found in the ground state at finite magnetic field persists, in modified form, into low energy transport states in the absence of the magnetic field. This indicates that such current loops are intrinsic to the twist bilayer in situations of broken time reversal symmetry, and is consistent with the independence of the lattice parameter of the equilibrium current loop lattice from the magnetic length. In a low energy picture the twist bilayer can be described as two Dirac-Weyl sub-systems coupled by a non-abelian interlayer moir\'e potential\cite{san12, shall16}. As this potential is known to drive an interlayer precession of quasi-particles\cite{san12}, we conclude that transport is governed by an interplay of the applied bias potential and the moir\'e potential which, in the low energy transport states, manifests as ordered current loop lattices.

In the presence of an out-of-plane magnetic field similar current loop structures can be observed in transport states, see Fig.~4(c-d), with, interestingly, those that block longitudinal transport showing a similar interlayer current modulation to that found in the ground state and those that transmit a structure closer to the $B=0$ current modulation. In fact, in the blocked state the current flow close to the edge of the gated layer is, due to the current loop array, against the direction of the applied bias, see Fig.~4c. However, one should note that the width of the ribbon here is only $\sim 4 l_B$ and a ``bulk'' Landau level structure is not seen in this geometry.

{\bf Discussion.} We have uncovered imprints of the moir\'e in the current densities of both equilibrium and non-equilibrium states of the twist bilayer. In equilibrium the zero mode Landau level exhibits an ordered array of permanent current loops throughout the bulk, consisting of a ``convection cell'' current configuration attached to each moir\'e spot of the lattice. Such current loop lattices, in modified form, are also found in transport states near the Dirac point, as seen in the low energy scattering states of a graphene nanoribbon junction overlaid by a twisted layer. Remarkably, here the current loop array orders such that current in the second layer -- which has a magnitude of $\sim50$\% the current strength in the gated layer -- flows against the transport direction. Loop lattice transport states are found for both zero and finite magnetic field, showing the phenomena to be intrinsic to the twist bilayer in situations of broken time reversal symmetry. The physics of the equilibrium loop lattice is therefore underpinned by the interlayer precession property of the moir\'e potential\cite{san12}, rather than the physics of a confining potential of length scale comparable to the magnetic length which drives bulk currents in 2DEGs at rough semi-conductor interfaces\cite{vig94,vig95}. While closed current loops in equilibrium are likely to be difficult to observe in experiment, although they may interact in interesting ways with ordered arrays of intercalated impurities for which the moire forms a natural geometry\cite{sym15}, the persistence of such current loops in transport states shows that this phenomena should be observable by local probes such as scanning gate microscopy.

{\bf Method and computational details.} In order to study the twist bilayer in an out-of-plane magnetic field we employ a nanoribbon geometry that allows free choice of the magnetic field strength, as well as allowing for the study of edge currents. For all systems studied we ensure that the nanoribbon width is at least an order of magnitude larger than the magnetic length in order to achieve a good bulk. The large system sizes that small angle twist bilayers nanoribbons then inevitably entail - our nanoribbon unit cells typically contain up to 40,000 carbon atoms - necessitates the use of a semi-empirical tight-binding method. We will use the tight-binding method of Ref.~\cite{lan13} which was deployed in that work for the study of twist bilayer flakes; it consists of the environment dependent method of Tang et al.\cite{tan96}, but re-parameterized by performing a least squares fit to the high symmetry eigenvalues from a number of small unit cell few layer graphene systems generated \emph{ab-initio}; for details we refer the reader to Ref.~\cite{lan13}.

We introduce an external magnetic field into our calculations via the standard Peierls substitution\cite{pei33} as an additional phase in the two-center hopping integrals:

\begin{eqnarray} 
  t_{ij} \rightarrow \text{ } t_{ij} \text{ exp}\left\{i \frac{e}{\hbar}
  \int_{\mathbf{r}_i}^{\mathbf{r}_j}{\mathbf{A} \cdot d\mathbf{r}} \right\},
\end{eqnarray}
with a Landau gauge $\mathbf{A} = \left(0,Bx,0 \right)$ such that $\mathbf{\nabla} \times \mathbf{A} = B \mathbf{\hat{z}}$ (the graphene plane taken to lie in the $xy$ Cartesian plane).

For calculating the electron and hole currents we follow the derivation of T. Todorov\cite{tod02} and determine the current from atom $n'$ to atom $n$ as:

\begin{equation}
I_{ij} = \frac{2e}{\hbar}
\Im \sum_{\gamma \gamma^\prime} H_{j \gamma,\, i \gamma^\prime} \, c_{i \gamma^\prime} c^\star_{j
\gamma},
\end{equation}
where $H$ is the real space tight-binding Hamiltonian, $c_{i\gamma}$ are the expansion coefficients of the eigenvectors in the tight-binding basis, and $\gamma$ is the orbital index. The current at each site in the lattice is found from

\begin{equation}
  \mathbf{j}\left(\mathbf{R}_i\right) =
  \sum_j I_{ij} \hat{\mathbf{r}}_{ij}
 \label{eqcur}
\end{equation}
with $\hat{\mathbf{r}}_{ij}$ the unit vector pointing from site $i$ to site $j$.

The problem of finding a commensuration cell of the twist bilayer leads to a discrete set of unit cells labeled by two integers, which we will refer to as $p$ and $q$ following the notation of Refs.~\cite{shall10,shall13,shall16}. For a nanoribbon geometry, these unit cells evidently define the edge structure which, as might be expected, is always a low index facet. For $p=1$ nanoribbons the edge comprises one zigzag segment per nanoribbon cell edge, with the remainder of the edge having the armchair type termination.

For the transport calculations presented in Fig.~\ref{transfig}, we have simplified the tight-binding model to the two-center hopping function

\begin{equation}
V\left(\mathbf{R}_i-\mathbf{R}_j\right) = A \text{exp}\left(-B \left\vert \mathbf{R}_i - \mathbf{R}_j                    \right\vert^2\right) \, ,
\end{equation}
and constructed the geometry and Hamiltonian using a code specifically written for this purpose. We have then converted them to the format of the TBtrans code \cite{Papior2017}, which was used to calculate the transmissions and orbital currents using Green's function formalism.  The model parameters have been chosen such that they reproduce the DOSs for AA- and AB-stacked bilayer graphene as given in Ref.~\cite{Rozhkov2016}. We have then made use of the SISL code \cite{zerothi_sisl} to calculate the bond currents $I_{ij}$ from the lattice site $\mathbf{R}_j$ to $\mathbf{R}_i$. From the bond currents, we have calculated a vector field \emph{via} Eq.~\eqref{eqcur}, which has then been coarse grained in order to obtain the representation shown in Fig.~\ref{transfig}.  

{\bf Data availability.} The data that support the findings of this study are available from the corresponding author upon request.

{\bf Author contributions.} The project was framed by Shallcross and Meded, calculations were performed by Landgraf, Fleischmann, and Weckbecker, using a code written by Weckbecker and Shallcross. All authors contributed to the interpretation of results and writing of the manuscript.  

{\bf Competing financial interests.} The authors declare no competing financial interests.

{\bf Acknowledgments.} This work was supported by the Collaborative Research Center SFB 953 of the Deutsche Forschungsgemeinschaft (DFG).

\bibliography{paper.NR}

\end{document}